# An Overview of Enhancing Distance Learning Through Emerging Augmented and Virtual Reality Technologies

Elizabeth Childs[o*] Ferzam Mohammad[o†] Logan Stevens[o‡] Hugo Burbelo[§] Amanuel Awoke[¶]
Nicholas Rewkowski[‖] Dinesh Manocha[**]

University of Maryland
[o]These authors contributed equally to this work

## Abstract

Although distance learning presents a number of interesting educational advantages as compared to in-person instruction, it is not without its downsides. We first assess the educational challenges presented by distance learning as a whole and identify 4 main challenges that distance learning currently presents as compared to in-person instruction: the lack of social interaction, reduced student engagement and focus, reduced comprehension and information retention, and the lack of flexible and customizable instructor resources. After assessing each of these challenges in-depth, we examine how AR/VR technologies might serve to address each challenge along with their current shortcomings, and finally outline the further research that is required to fully understand the potential of AR/VR technologies as they apply to distance learning.

## 1 Introduction

Augmented reality (AR) is an emerging form of digital experience in which the user's visual perception of the world around them is augmented using a computer-generated graphical overlay. This overlay is superimposed onto the user's view through a combination of sensors and algorithms, enabling virtual components to be projected into the real world, sometimes on the lens of a head-mounted display (HMD) that the user wears. In addition to AR, Virtual Reality (VR) is a related technology that also leverages computer technology in order to create an artificially rendered digital experience, but crucially differs from AR in that the visual experience of VR is entirely computer-generated, while AR superimposes digital rendering on top of the real world.

While the technical feasibility of AR and VR has existed for decades, AR and VR have only recently become commercially accessible due to the advent of mobile platforms (e.g. iOS and Android) combined with rapid advances in consumer grade hardware [93]. Applications of AR and VR have the potential to benefit many fields, but one of the most impactful applications for these technologies is leveraging them in an educational context [10, 86].

Being able to support learning with the addition of immersive virtual environments seems promising; due to the fact that education as a whole has been disrupted recently due to the COVID-19 pandemic, the importance of evaluating exactly how AR and VR technologies can improve the learning experience cannot be overstated. Distance learning, a method of studying in which lectures are broadcast or classes are conducted by correspondence over the internet without the students needing to physically attend a school or college, has also already become an integral part of many schools around the globe prior to the pandemic [23]. Greenberg defines contemporary distance learning as "a planned teaching/learning experience that uses a wide spectrum of technologies to reach learners at a distance and is designed to encourage learner interaction and certification of learning" [34]. Distance learning is a particularly promising educational application of AR and VR due to its unique pedagogical structure and rapidly expanding technological dependency. AR and VR have the potential to create unique learning opportunities that allow course content to be taught and presented in ways that may have otherwise been extremely challenging if not impossible. This is of value in any form of education, but could vastly improve distance education because of the issues students regularly face in distance learning environments.

There is considerable research on the use of AR and VR in education. Most of these systems or investigations were based on earlier hardware prototypes that were not easy to use, expensive or not widely available. As a result, the overall applications or usage of VR and AR technologies for education and distance learning have been limited. Recent availability of low-cost head mounted displays and increasing software support has motivated researchers and developers to use these technologies for other applications. At the same time, there is lack of useful data in terms of using AR and VR for distance learning. Our goal in this survey is to bring focus to this gap in data, and to deliver a clear overview and survey analysis to the current state of the art in the overlap of AR and VR in distance learning.

In this paper, we first assess the educational challenges presented by distance learning as a whole, and then identify 4 main challenges that distance learning currently presents as compared to in-person instruction: the lack of social interaction, reduced student engagement and focus, reduced comprehension and information retention, and the lack of customizable instructor resources. After assessing each of these challenges in-depth, we examine how AR and VR interaction mechanisms, and visual immersion might serve to address each challenge along with their current shortcomings. Finally, we outline the further research or development which is necessary to fully realize the potential of AR and VR technologies as they apply to distance learning. This survey is intended to provide a comprehensive overview of prior work, the current state of the art, and highlight many open research areas with respect to use of VR/AR/XR technologies in distance learning. We assume that the reader has some exposure to basic concepts in VR and AR. Examination of the provided tables is strongly encouraged for further clarification on the information provided in this survey.

[*]e-mail:ehchilds@terpmail.umd.edu
[†]e-mail:fmoham18@terpmail.umd.edu
[‡]e-mail:lsteven7@terpmail.umd.edu
[§]e-mail:hburbelo@terpmail.umd.edu
[¶]e-mail: aawoke@terpmail.umd.edu
[‖]e-mail:nick1@umd.edu
[**]e-mail:dmanocha@umd.edu

## 2 IMMERSIVE ENVIRONMENTS

This section provides the background and definitions planned for use throughout the rest of this paper. Immersive environments blend a digital, computer generated world with the real world. They augment, or fully replace a user's perception of the spatial world around him or her, placing the user in an altered environment. The ratio of real objects to digital objects in a user's view affects the amount of digital augmentation, and was defined in Milgram's Mixed Reality (MR) continuum [61]. In the continuum, a real to digital ratio of 1:0 would be the real world at the beginning of the spectrum. As digital content is added, computer generated images are superimposed onto the user's view. The amount of digital content added to the user's view determines the program's position on the MR continuum. For example, simple text or 2D images that appear in the user's view (such as when Google Maps AR gives an arrow and street name for walking directions [33]), would be at the lower end (left) of the continuum. Increasing the immersion, such as adding a realistic 3D couch in your living room before buying it [11] moves the application further right on the MR spectrum. Environments that are entirely computer generated but allow the user view of certain real elements, such as a hand or select real objects, would be considered augmented virtuality. With this definition, everything with a real to digital ratio greater than 0 and less than infinity constitutes mixed reality, with augmented reality being at the lower end of the spectrum and augmented virtuality being at the higher end.

It should be noted that over time, the definitions of AR and MR have evolved since Milgram first defined the MR continuum. Conventionally, anything that is not purely real or purely virtual has been defined as augmented reality. Technically, Microsoft defines their HoloLens as mixed reality [59] to distinctly differentiate it from lower resolution AR glasses, such as the raptor [26], which fits Milgram's true definition of augmented reality. It should be noted that for the purposes of this study, all immersive experiences except for pure virtual reality will be addressed as augmented reality, both to match the convention of similar studies and for simplicity of understanding.

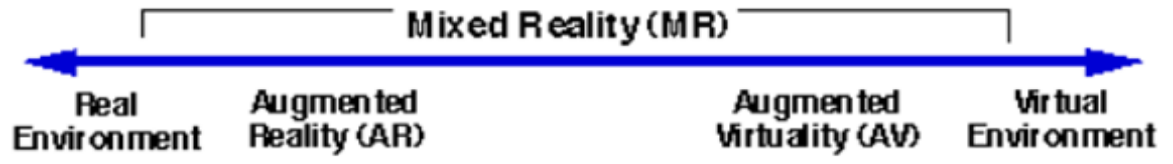

Figure 1: *Milgram's MR Continuum*. [61]

When the user's view is completely virtual (real to digital ratio of 0:1), the experience is considered VR. With VR, the user has no view of the real world and can only see a computer generated world that exists in the same spacial area as the real world. For the purpose of this paper, AR will be considered all devices in the middle of the MR continuum, where both digital and real images are visible, with VR being entirely virtual environments. We will consider the hand tracking and body tracking that many present day VR headsets have as completely virtual for the scope of this research, as this often appears within a VR environment as a rendered image based on sensors.

### 2.1 Devices for Immersive Environments

In addition to the blend of real and digital content that defines immersive environments, they can also be categorized by the way a user experiences the digital content. Immersive Devices have control over the user's entire point of view, whereas non-immerse devices only augment part of the user's view.

#### 2.1.1 Immersive Devices

Immersive devices are typically headsets, or head mounted displays (HMDs). They may include earpieces like noise-cancelling headphones to control the user's auditory field, and most HMDs incorporate handheld controllers which increase the methods of input for the immersive experience. With VR, the user actually feels transported into the new environment, and with AR, the digital information is cohesive across all of the user's visual field. In terms of education, immersive displays have been shown to increase memory recall over non-immersive displays [84] and help with information retainment.

However, immersive displays do have limitations. While inexpensive immersive displays exist, such as Google Cardboard [32], they do not have the graphics and computing ability to offer truly immersive environments. Devices like the Oculus Quest 2 have begun to enter the consumer market due to lower price tags than what VR headsets initially launched at (i.e. as of June 14th 2021, one can buy a 64 GB Oculus Quest 2 off of the Oculus website for $299 USD. However, when the Oculus Rift launched it was priced at $599 USD). That said, the more immersive headsets like the Oculus Quest 2 may still be too expensive to bring into educational environments.

#### 2.1.2 Non-Immersive Devices

Non-Immersive devices include handheld displays, and typically use a smartphone or tablet. The user can still move around in the virtual or augmented world; however, all of the digital content is viewed through the device screen. AR and VR applications can also tap into the device's camera, gyroscope, and/or accelerometer to enhance an immersive experience. Non-Immersive devices are used more prevalently for AR education.

Advantages of non-immersive devices stem mostly from their accessibility. While non-immersive devices may not have the same technical capability or depth of experience as immersive devices, over half of the world has a smartphone [3]. This makes non-immersive devices ideal as educational tools because they are accessible, inexpensive, and relevant for current educational populations.

For example, smartphones allow the user to experience MR by altering the screen content based on the spacial location of the device. The MR content can then be viewed using a handheld device by either looking through the phone or by strapping the phone into a head mounted display (HMD) and onto the face. Additionally, the device's accelerometer, gyroscope, microphone, and speaker can provide additional levels of immersion.

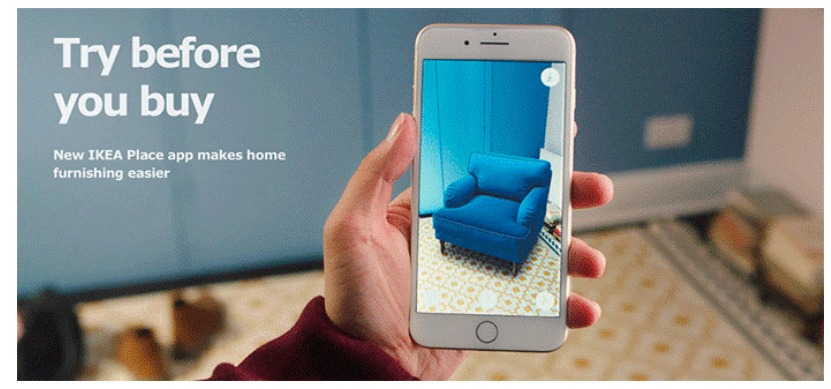

(a) Non-Immersive AR Device [5]

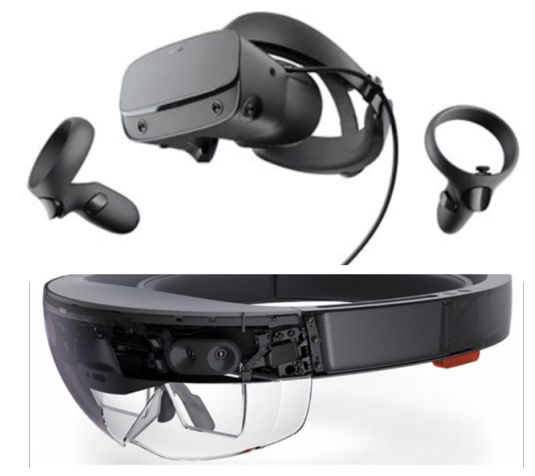

(b) Immersive AR (top) [4] and VR (bottom) Headsets [59]

Figure 2: *Immersive and Non-Immersive Devices*.
*Non-Immersive Device* (a) - See digital content through a mobile screen. *Immersive Device* (b) - View digital content through a headset.

## 3 EXISTING FORMS OF DISTANCE EDUCATION MODELS

Online learning had become a critical tool utilized by close to 70% of higher education institutions prior to the COVID-19 pandemic [23]. Instructors were already making use of virtual educational strategies such as asynchronous course delivery, synchronous online course delivery, or a flipped classroom model that utilizes both. Understanding these existing forms of online education and the benefits and drawbacks of each system will highlight how AR or VR can address the issues in these systems.

### 3.1 Asynchronous Learning

An asynchronous learning model involves students interacting with and responding to material at their own pace. Though students will still have to meet deadlines, the timeline they follow for working through material is flexible. An example of this model is a teacher posting all lecture material and class assignments with their respective deadlines at the beginning of a course. The students would then manage their time however they see fit in order to get these tasks done by the deadline. Asynchronous courses can use tools like video recordings of lectures, digitally distributed typed or written material, discussion posts, or a combination of these tools. This delivery method gives students flexibility in regards to when they access material, which can be crucial to academic success for non-traditional students navigating other time commitments like a job or taking care of their families [44].

### 3.2 Synchronous Learning

A synchronous model is what traditional face-to-face courses use, where students and their instructors meet regularly for teaching sessions, tests, and homework submissions. Synchronous online courses may require a different set of resources as compared to the asynchronous model. Learning Management Systems (LMSs) like Canvas or Blackboard have been adopted by the majority of universities as supplementary resources intended to support teaching and learning activities [64], and different features of LMSs can be helpful for various learning models. Certain LMSs offer tools that allow for synchronous online course delivery, like Blackboard's "Blackboard Collaborate" (formally Elluminate Live! [16]). Universities have also adopted other third party options to conduct online classes such as video conferencing software, e.g. Zoom [6]. Tools commonly found within such platforms include video and text chat features, a participant list, polling/surveying tools, a raise hand tool, and the ability to record a session [7]. Synchronous courses can also make use of "screen-share" features that allow students to see what is on an instructor's computer screen, and these can be used in conjunction with digital content like PowerPoint slides to help relay course material.

### 3.3 Flipped Classroom Model

A "flipped classroom" is a teaching model where "that which is traditionally done in class is now done at home, and that which is traditionally done as homework is now completed in class" [13]. This typically involves students viewing "lecture" material, or the material related to the introduction of concepts, at home as opposed to in the classroom. Class time is then spent on "more enriching activities" [74] that allow students to deeply explore concepts they learned from the lecture material. In an all-online environment, a flipped classroom could be implemented using asynchronous methods to distribute course material prior to synchronous sessions. The synchronous sessions would then focus on student application of knowledge learned from their material. This would involve the resources required of both asynchronous and synchronous teaching.

### 3.4 Current Distance Learning Technologies: Limitations

There are clear benefits in regards to existing distance learning education models, but several drawbacks have been noted as well. Distance learning in any form appears to decrease opportunities for social interaction and discourse as compared to traditional, in-person courses [23]. Presently-used digital formats have also been seen to disengage students from their courses due to technical issues or unsatisfactory experiences with the online class setting [57]. Schools that only offer online courses have unfortunately seen higher dropout rates than those which are in-person [78].

These issues indicate that the way we currently deliver distance education could be damaging students' learning opportunities as compared to an in-person education. However, the immersive nature and virtual components of AR and VR can create opportunities to address current drawbacks of distance education. We explore the different ways AR and VR address these issues throughout section 4.

### 3.5 Distance Learning and Gamification

Recent years have shown developments in the design of online coursework, and these advancements are worth understanding when considering how we can effectively make AR or VR useful for the classroom. Concepts like gamification have been applied to digital educational resources as a way to improve many aspects of the learning process. Using "game mechanics" like points, levels, challenges, or badges in educational activities has helped improved the ability to learn new skills by 40% compared to without the use of those mechanics [31]. Gamification has also led to increased commitment and motivation when participating in activities beyond just education [48]. Effective gamification requires the consideration of many aspects of a course like learning objectives or the characteristics of learners [48], and successful implementation of gamification in a curriculum parallels many aspects of what "good" teaching already looks like [81]. Educational tools which make use of effective gamification are seeing great popularity. Duolingo, a phone app designed to help people learn languages through "gamified" lessons, had around 40 million monthly active users as of December 2020 [17]. Kahoot!, an online game where you can make your own quiz competitions on any topic of your choosing, reached 5 billion cumulative players around the world as of January 2021 [36]. Gamifying education has seen great support as mentioned previously; learning to leverage gamification both in traditional distance learning models as well as when integrating immersive technologies into education could have a strong positive on educational systems if done correctly.

## 4 CHALLENGES IN DISTANCE LEARNING

In this section we explore the specific issues faced by students utilizing current distance education methods. After describing an issue, we explore existing work in how AR and VR are being used to address the issue being faced. Finally, we consider what work remains to totally resolve the issue when using immersive technologies.

### 4.1 Social Interaction

We will define social interaction as "an intentional event in which one person's behavior is directed toward another person or is in response to the other person's behavior" [30]. This definition remains the same both in non-digital settings (i.e. a setting that does not use AR or VR explicitly to change the user's experience) as well as digital, though the interactions may appear differently between different media. For example, social interaction in a digital setting may be a student sending a message via text chat to another student, while in non-digital settings like in-person classrooms students can talk directly to one-another. Virtual classrooms, such as Blackboard Collaborate, are platforms which allow students to interact with the instructor or other students through text or voice chat. These platforms can also enable students to see course material, like slides, remotely. Virtual classrooms and LMSs are different systems. A LMS provides features needed to handle the logistics of the entirety of a course, like the ability to post class-wide lesson materials, assessments, and feedback for everyone to see. Virtual classrooms attempt to replace the physical space of a classroom by focusing on creating opportunities for real-time interaction between users or users and content. Virtual classrooms can also differ from AR or VR learning modules in several ways. Existing virtual classrooms are designed to entirely replace the classroom for the duration of a whole course as opposed to being a substitute for single lessons. AR or VR learning modules more commonly replace specific, single lessons (e.g. the virtual classroom Blackboard Collaborate provides the tools needed for students and teachers to engage with each other as well as material during any digital class period, while a VR learning module for a course would have to be predefined to support the content related to a specific lesson). Existing non-immersive virtual classrooms are typically web applications and only encourage as much social interaction as web-based voice/video platforms can. While distance learning methods give students more flexibility in accessing course content, studies have shown that students can feel "disconnected", or struggle to practice social interaction with others in these online environments [44, 57]. Traditional face-to-face education is shown to be "more likely to promote collaborative learning, student-faculty interaction, effective teaching practices, quality of interactions, and discussion with diverse others" when compared to a virtual classroom [23]. Quality social interactions have previously been claimed as a way for young children to develop cognitively [12, 18], so the absence of these interactions can prevent this cognitive growth [19, 68]. This means students in current distance learning models are losing aspects of their education due to the lack of social engagement.

#### 4.1.1 AR Research to Address Social Interaction

Since distance learning can decrease social interaction, research has used AR to facilitate social collaboration when users are separated. Attempts to use computer vision to recreate user movements into remote avatars have been under investigation for several years now [58, 73]. Raskar had previously shown that this could be done by using high depth cameras to capture a user's 3D image which could then be projected remotely for other users. However, since this method required high depth camera images from multiple perspectives, it was not practical for the average learning environment. As a result, recent researchers have turned to other methods to enhance collaboration like generating avatars using low-cost cameras [58] or projecting 2D images [14]. For example, Michael et al. explored ways to cut costs for avatar generation using machine learning. He started by training two machine learning models on a dataset of 3D body scans. One of these models was trained to identify and extract the parts of the image which were the body, and the other was trained to then apply those extracted images onto a 3D model. After training both these models, a low-cost camera could be used to get full-body images of a person from multiple perspectives, and then the machine learning models could be fed these images to generate a 3D avatar. Michael et al. did use one other high-resolution scan of the face so that facial features were captured, but this was more cost-effective than previous, similar approaches which used several high-end cameras to get the input images [58].

Another example of AR being used to increase social interaction digitally was seen when Billinghust et al. used HMDs and AR marker cards to project images of participants for video conferencing of physical cards [14]. Billinghurst et al. had participants wear an HMD and carry an AR marker card which projected video feed of their partner if seen through the HMD. The partners would then walk through and participate in an art sale. Billinghust compared the AR conferencing method to audio only conferencing and traditional video conferencing under the same context. In the AR conference, participants could move freely around the room with their cards and see their partner. In contrast, the traditional video conferencing took place over a desktop computer, while the audio conference was a phone call. Participants were then asked to rank their sense of presence as well as the conference's similarity to in-person meetings. Notably, the AR conference scored significantly higher (p value $<.05$) than audio or traditional video conferencing. The greatest drawback of the AR group was that the HMD obscured participants faces and field of view, which made communication harder during the AR conference.

| Study | Device | Result |
|---|---|---|
| Raskar [73] | Projectors and Surfaces | Details theoretical "office of the future" that could connect collaborative spaces in different locations using projectors and wall surfaces |
| Billinghurst [14] | Head Mounted Display | AR conferencing can provide an increased sense of presence and improve transmission of communication cues for remote users |
| Pejoska [67] | Mobile | Social AR successfully used to enhance video calls with overlaid drawings |

Table 1: *Results of AR Studies to Address Social Interaction.* **1st Column:** Reference. **2nd Column:** AR device used for the study. **3rd Column:** Result of AR Study. These results indicate AR can be useful in creating creating social interactions for education.

Pejoska et al. used AR to aid social interaction while teaching construction workers [67]. They designed Social AR, which helped the workers communicate over large distances while on the job. Social AR allowed users to send voice and text messages through their mobile phone, as well as augment their video for others to view. For example, a user at one end of a construction site would be able to show their video feed in real time, annotate it, and the annotations would also appear to other users during creation. This allows workers to be able to see what their coworkers are seeing. Additionally, text features allow participants to discuss the annotations and video feed as if they were interacting in person. This research shows the potential of AR to increase social interaction during information sharing, as well as suggests methods for social interaction to be incorporated into distance learning via AR.

These studies indicate that AR can be used to provide a sense of presence with other classmates, which is crucial for distance learning. This makes AR a great avenue for enhancing distance education, as it enhances video calls and connects students collaboratively to bring social interaction in distance learning closer to that of traditional education.

#### 4.1.2 VR Research to Address Social Interaction

The ability to socialize despite being physically separate is a critical aspect of the efficacy of virtual reality experiences. One significant reason VR is viewed as a unique alternative to common virtual communication tools like text or voice chat is its ability to more effectively emulate human interaction relative to conventional video conference calls (such as Zoom). Utilizing data from the National Survey of Student Engagement, Dumford and Miller analyzed first, fourth, and fifth year college students and found a significant correlation between online classes and social interaction [23]. These students had lower levels of collaborative learning, less student-faculty interactions, discussions, and lower quality of interactions [23]. VR can aid in this scenario because it will provide a simulation of face to face interaction while still maintaining some of the convenience of online learning.

Socialization and interaction with VR appears to match in-person interaction experiences closely. In a study that focused on social interaction quality in VR, researchers examined how the reduced social information and behavioral channels in immersive virtual environments compared to that of the real world. They also measured whether working in virtual reality environments as a whole were a viable alternative to the real world in terms of task completion and socialization [75]. Roth et al. had participants interact in communicative scenarios (a negotiation) both through a virtual environment as well as a real world environment. After comparing the effectiveness in communication, socialization, and overall task completion across two different environments, one virtual reality-based and one in the real world, the researchers were able to confirm that "differences in effectiveness in the communicative role play were not present." This means that the VR experience was equally as effective in enabling participants to adequately socialize and communicate while attempting to complete a task.

Social Virtual Reality (SVR) has recently become an area of research which exemplifies the use of VR in order to increase the feeling of presence and engagement under virtual environments. SVR is defined as 3D spaces where several users are able to interact with each other through VR apparatus like HMDs [55]. SVR is still a newer area of research, which warrants greater investigation in the subject itself. However, the existing findings about the positive effects of SVR in educational environments on students' feelings of presence and engagement when compared to 2D environments with the same objectives indicate that SVR and its related technologies are useful in building socially engaging virtual environments with VR [65, 66]. We have already seen studies showing that SVR produces emotional responses similarly to what users would experience in person during social interactions, and that even "social group dynamics" transfer to SVR settings when they exist offline [63]. When used as a tool for safety courses related to work on construction sites, "collaborative VR" was found to "improve safety experiential learning, the identification of potential job hazards as well as worker's risk cognition," [51]. Being able to continue to regularly use the SVR system also allows learners to safely continue practicing construction-related skills or activities so that they would be better prepared to prevent accidents in the field [51]. Even forms of non-verbal communication like body language or gesturing, which are characteristic to in-person social interaction, are used in SVR and "simulate experiences that are similar to offline face-to-face interactions," [55].

SVR in tandem with gamification is an emerging research area which promises to address some of distance learning's more problematic issues (i.e. poor engagement, absence of presence, and inability to interact with others). The work of Mystakidis touches on some of these topics by exploring the effect of gamified curriculum design on student engagement and critical discourse using VR [65]. That work compared qualitative findings from a non-gamified section of a post-graduate program on education in VR to a gamified section of the same program, where several game mechanics were introduced to the curriculum (e.g. story was introduced to learning objectives, students could level up by completing tasks, there were a variety of quests available for students to choose from based on their interests and skills, etc.). Both sections of the course had many VR elements (e.g. virtual worlds, social interaction in virtual settings, 3D avatars for characters, etc.). Students from the non-gamified section of the program "demonstrated predominantly structure-dependent engagement", where students completed tasks they were assigned and were able to utilize exploratory tools like AI-bots "fairly effectively". However, students under the gamified section of the program appeared to be more "playful" when engaging with the environment and course content, demonstrating feelings of "enjoyment and relaxation". Students from the gamified section were also more experimental with the AI-tools they had access to, and they had even been seen attempting to brainstorm ways the AI-tools could be used to achieve learning objectives in courses they might teach in the future. Mystakidis found students subjected to the gamified distance learning were noticeably more engaged and self-directed as compared to students in the non-gamified program, and that "gamified elements elicited students' interest, motivation, and autonomy towards critical engagement." [65]

Monahan et al. explored how VR can improve social interaction amongst peers in school compared to e-learning without VR using their tool CLEV-R. CLEV-R is a collaborative VR environment made to emulate a university setting and it provides students with collaborative tools as well as designated virtual spaces for social interaction with others. 89% of users in this experiment "felt the environment was an effective means of social interaction," and educators who tested the system claimed they wanted to see it used as an educational tool in the future [62]. CLEV-R enabled people to engage and interact more effectively, demonstrating how VR can be used to help with overcoming the lack of social interaction present in today's distance learning systems.

Studies in VR social interaction suggest that advanced computer graphics such as avatar facial expressions may not be necessary for VR communication [75]. This can make VR education applications more affordable and easier to develop and adapt for educational institutions.

| Study | Device | Result |
|---|---|---|
| Monahan [62] | Head Mounted Display | Participants interacted successfully and were engaged with the VR environment |
| Roth [75] | Head Mounted Display | Absence of cues like gaze or facial expression do not change effectiveness of communication in VR |
| Southgate [79] | Head Mounted Display | Embodiment helped different students engage organically |

Table 2: *Results of VR Studies to Address Social Interaction.* **1st Column:** Reference. **2nd Column:** AR device used for the study. **3rd Column:** Result of VR Study. These results indicate complex gesture algorithms and graphics are not necessary for educational VR social interaction.

#### 4.1.3 Future Research in Social Interaction

While current research in AR and VR has developed additional opportunities for social interaction in distance learning, there are still limitations to the state-of-the-art. There are currently no standard metrics used for measuring social interaction in AR or VR, so the field would greatly benefit from establishing a standard to determine what MR technologies and methods can best help in social interaction. In terms of hardware, HMDs have a limited field of view and obstruct users' faces. As a result, it is harder to make eye contact and experience a social connection when both users have a headset [14]. Future research into facial communication with HMDs (such as less obtrusive hardware, or another way to communicate non-verbally) can enhance social communication when using HMDs. In VR, more educational tools are becoming publicly available as well [62]. It could be valuable to evaluate these tools and how effective they are in both encouraging discourse as well as facilitating lessons. Understanding what makes these environments effective or ineffective would help us build better VR or AR educational platforms. This information could also help us see how to modify traditional distance learning education methods so that they are more effective in encouraging discourse.

When networked properly, VR embodiment can also augment social engagement within education. Southgate utilized VR social engagement for teaching high school and elementary students [79]. Particularly, VR effected how students interacted together. Personal appearances and aspects of social presentation can be adjusted at will, so that many potentially-divisive personal differences can effectively disappear. For example, a student's special-needs differences were masked by his knowledge and presentation within VR with the other students [79]. Future research is needed on investigating the mental effects of being able to portray yourself as an avatar in educational environments. This would be beneficial so as not to perpetuate existing issues in problematically masking certain physical characteristics, such as race, or mental characteristics, such as neurodivergence.

### 4.2 Student Engagement and Focus

Asynchronous and synchronous courses have both faced criticism in regards to how engaging they are for students. In asynchronous classrooms, students entirely lose the non-verbal communication present in face-to-face communication. This can damage students' sense of social presence and make it harder to stay engaged in a class [44]. In a study by McBrien et al. on how synchronous online courses affect student engagement, 9% of students in the study felt "disconnected" as a result of the digital course format. One of the reasons for feeling "disconnected" is the minimal non-verbal communication available. Technical issues (i.e. poor internet connection or trouble with joining into a class) can also have a detrimental effect on how "connected" students feel with a course. The negative responses from students in the study by McBrien et al. included comments like "sometimes it was hard to keep up with the messages, listening to commentators, and reading over the lessons," "lack of visual stimulation during lecture," "frustration signing on and getting kicked off," and "microphone troubles." These comments seem to indicate that certain students would have preferred to have fewer stimuli going on during courses. Conversely, there was a claim that there was too little visual stimulation, suggesting that the material being focused on by the teacher may have been too bland given the online setting. The last couple of claims indicate the significance of technical issues in the educational experience; any trouble with getting the class to work as it is expected to seems to greatly take away from the learning experience [57]. John Keller's ARCS model for motivation claims that motivation stems from four major components: attention, relevance, confidence, and satisfaction [45]. Looking at the students' claims from the perspective of Keller's ARCS model, it seems like the different issues behind the complaints could have made it harder for students to maintain uninterrupted attention or feel satisfied with the online course. Decreased satisfaction or attention has a negative effect on a person's motivation according to Keller, and decreased motivation may have damaged students' abilities to engage with the class.

#### 4.2.1 AR Research to Address Student Engagement and Focus

Researchers have already turned to AR to investigate its effect on engagement and focus. Ferrer-Torregrosa et al. utilized AR to teach anatomy students in a flipped classroom distance learning model [27]. Participants were given AR books, traditional notes with pictures, and videos to learn anatomy content. They were then assessed by which tool (book, notes, or video) kept the students more engaged and focused. Notably, the AR books were the superior method for holding a student's attention and focus without the teacher present.

Additionally, consumer mobile augmented reality applications have been assessed for their usage in increasing student motivation. Anatomy 4D is a free mobile AR application for teaching students anatomy. The software shows a 3D human body with annotations labeling the different body parts. This software can be downloaded and used without a teacher present, and offers a great example of AR remote learning tools. Through pre-usage and post-usage tests, Khan et al. determined a significant increase in student motivation from using Anatomy 4D [47].

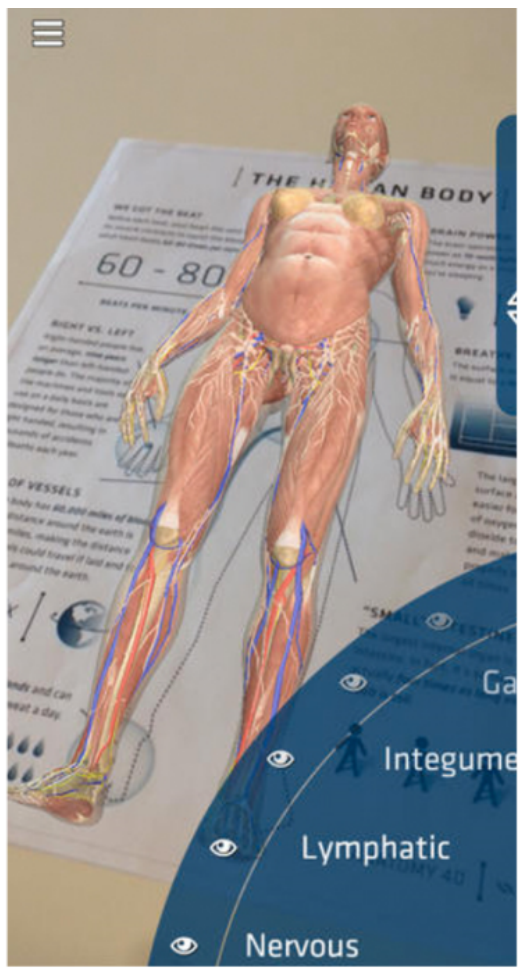

Figure 3: *Preview of Anatomy4D Application* [47]

Although not explicitly tested within distance learning, self-embodiment within AR can also contribute to engagement. Johnson and Sun utilized Spatial Augmented Reality (SAR), in which parts of the anatomy are projected onto ones' bodies. Using a Microsoft Kinect and projector, Johnson and Sun were able to provide size-appropriate anatomy spatially for the students. Within a K-12 environment, both teachers and students found this more engaging than a typical lecture [43]. Hoang et al. had conducted a study surrounding the same concept of overlaying anatomy, but also animated the display according to each participant, such as a beating heart rate [38]. This resulted in even more engagement, as students felt more connected to the display. Both were considered useful in schools, and Johnson's study saw an increase in learning anatomy in high school and tertiary education [43]. While this may be infeasible for many current distance education environments (as every institution/family/child may not have access to a projector and Kinect due to hardware costs), it is not improbable for future distance education environments to make use of SAR in distance education for increased engagement.

We have also seen AR have positive effects on long-term retention of concepts. Fidan et al. explored how AR applications affected students who were studying physics through Problem Based Learning (PBL) strategies. They found that AR "contributed to students' long-term retention of the concepts," and the students even emphasized that "AR applications were more useful, realistic, and interesting for their learning." This study also specifically took steps to remove the novelty of the AR device by providing students with opportunities to familiarize themselves with the devices for two weeks prior to the experimental process [28].

These studies show that AR had a significant increase in student engagement and focus, which were major difficulties faced in distance education. Several of these studies were shorter-term experiments. However, seeing long-term studies like Fidan et al.'s [28] which demonstrate similar benefits as those found in shorter term studies supports the idea that the benefits we are seeing students receive in short-term experiments may also be present in the long-term.

| Study | Device | Result |
|---|---|---|
| Fidan [28] | AR HMD | long-term retention of the concepts was seen to be improved |
| Khan [47] | Mobile | Motivation and attention was increased by 14% and 31% respectively |
| Johnson [43] | Projector | AR anatomy considered engaging by both students and teachers |
| Hoang [38] | Projector | Animated AR anatomy giving self embodiment |

Table 3: *Results of AR Studies to Address Student Engagement and Focus*. **1st Column:** Reference. **2nd Column:** AR device used for the study. **3rd Column:** Result of AR Study. These results indicate AR can elevate student engagement and focus.

#### 4.2.2 VR Research to Address Student Engagement and Focus

The advantages of using VR to teach educational objectives and improve student engagement are similar in many ways to the advantages of using a computer or interactive simulation, particularly a 3D computer simulation. Computer simulations create a sense of immersion by walking a user through the experience from a first-person perspective, such as a virtual tour of a historic building. Computer-based simulations have been used for many years in computer-assisted instruction (CAI). Steinberg asserts that "simulations make it possible to explore new domains, make predictions, design experiments, and interpret results" [80]. VR also gives the user a first person perspective of the experience; however, instead of interacting with the experience via a keyboard and mouse, the user can physically move around and utilize their hands to create a greater sense of immersion. As a result, VR maintains and expands on many of the same qualities and advantages of computer simulations listed above. Through these studies, virtual reality experiences have provided a unique ability to improve user engagement and focus due to the novelty of the interaction device and unique characteristics of the experience.

#### 4.2.3 Future Research in Student Engagement and Focus

VR and AR have already proven to be engaging supplemental resources for different subjects whenever available. There have also been reports that VR learning environments have shown an increase in positive emotions along with a decrease in negative emotions compared to both video and textbook options [8].

Going back to the study by McBrien et al., one of the issues mentioned was the lack of visual stimuli in the traditional online learning environment. Both AR and VR do a phenomenal job in creating engaging, informative environments, which should resolve this issue. However, another noteworthy comment made was the potential for too much stimuli in these virtual environments. VR or AR could easily involve more stimuli that the traditional distance education methods, so this seems like a factor to watch out for. Understanding when there is too much going on at once in existing distance education platforms could provide a basic idea of what a focused, effective teaching environment might look like in VR or AR. Research into what factors add to the educational value of a distance learning environment and which factors are excessive will help with removing unnecessary stimuli in these environments. That information could then be considered when designing VR or AR educational environments.

### 4.3 Comprehension and Information Retention

Schools that rely entirely on distance learning have previously held higher dropout rates than those which are in-person, but these statistics have also been tied to ineffective online teaching practices [78]. Simpson writes that "institutions have focused too much on the provision of teaching materials, especially online, and too little on motivating students to learn." Using VR in education has led to improved knowledge acquisition and understanding of material compared to those learning via video, and students learning via VR have shown better performance remembering than those learning via textbook or video [8]. AR technology has also been demonstrated to increase knowledge acquisition rate among a variety of educational subjects [93], and we have seen AR make abstract or complex concepts easier for students to visualize and understand. For example, AR could allow a learner to visualize phenomena like airflow or magnetic fields using virtual 3D objects [88]. Using immersive technology to illustrate these typically unobservable phenomena would give students a novel opportunity to understand new concepts within a familiar context.

#### 4.3.1 AR Research in Comprehension and Information Retention

AR technology has been demonstrated to increase students' knowledge acquisition rates in a variety of educational subjects [93] due to AR's ability to help visualize abstract and 3D objects. This is especially helpful for mechanical psycho-motor tasks, as AR overlays can show intuitive instructions for user activity without switching to an external screen. This was demonstrated with HMDs in Henderson and Feiner by using AR displays to facilitate assembly alignment [37]. Henderson and Feiner tested the time and accuracy for users to determine the location of parts, position them, and then align the parts together correctly. Participants aligned the mechanical parts almost twice as fast as compared to a traditional screen, with about 50% more accuracy. Additionally, subsequent studies showed an AR instruction to be almost equivalent to labeling each and every assembly part. However, since it can be unrealistic to try label every part, AR applications can be used as a valid learning substitute.

Non-immersive AR devices have also assisted in education when a teacher cannot be physically present. Schoop et al. used AR to teach construction for do-it-yourself projects [76]. Common household tools were converted to help with alignment, drilling holes, measuring wood, and etching details. The enhanced dynamic feedback exemplified AR's ability to provide clear instruction for individual tasks, as laypeople were able to use skilled machine tools and accomplish traditionally skilled machining tasks with the help of AR.

AR has also been researched in the academic teaching realm. In addition to studying motivation, Ferrer-Torregosa et al. analyzed how AR can increase students' abilities to self-learn, as well as their information comprehension [27]. When compared to traditional notes and videos, students felt more comfortable performing learning activities outside of class and increased their comprehension of the specified material after learning via AR. Ferrer-Torregosa et al. did not perform pre- and post-tests in order to confirm students increased comprehension of the material, but the students' perceived increases in comprehension (from metacognative survey analysis) can contribute to their actual understanding of material.

To increase information retention, Duenser et al. included haptics in their study of AR books [9]. Using a handheld device, participants were given books that introduced magnetism and electromagnetism through traditional or AR formats. All participants studied their respective materials for the same amount of time. Through pre-tests, post-tests, and retention tests (four weeks after the study), participants were evaluated both on their memory recall and comprehension of the physics concepts. The participants using the AR interactive book scored slightly higher in the post-test, especially for questions that required visualization (such as the left hand rule or drawing magnetic lines). However, both groups performed approximately the same in the retention test.

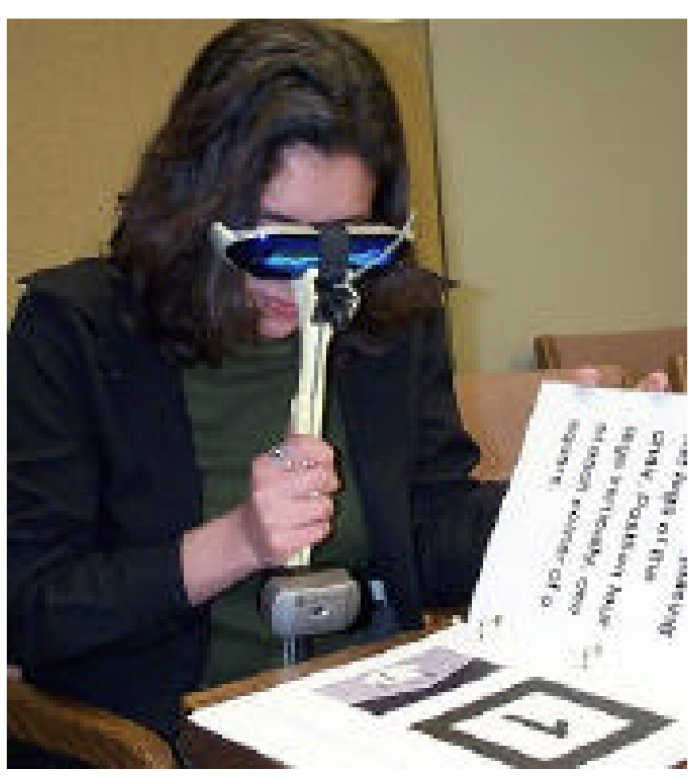

Figure 4: *Student using a handheld display to view an AR book* [15].

AR can also increase information retention by supplementing distance education. Küçük et al. created a mobile AR app to provide extra instruction in addition to lectures for anatomy students [50]. With the app, students could scan images of anatomy body parts and then view 3D animations and sounds to assist in learning. Students were able to retain more information using the AR mobile application while also facing a lesser cognitive load than with traditional studying. This represents AR's ability to aid in the difficulties of learning through distance education without interrupting the traditional distance education format. These studies indicate that AR can help students learn better, as well as more quickly compared to traditional methods. This makes AR a prime candidate for education, especially in distance education, where time with a teacher is significantly reduced.

| Study | Device | Result |
|---|---|---|
| Henderson and Feiner [37] | Head Mounted Display | Increase in learning and retention |
| Schoop et al. [76] | AR enhanced tools | Validation of AR to assist with learning in construction tasks |
| Ferrer-Torregosa et al. [27] | Handheld Display | Students learning with AR scored significantly higher compared to video or traditional learning |
| Dünser et al. [9] | Handheld Device | Students with AR scored higher than students learning with traditional books |
| Küçük et al. [50] | Mobile | Students retained more information with less cognitive load |

Table 4: *Results of AR Studies to Address Comprehension and Information Retention.* **1st Column:** Reference. **2nd Column:** AR device used for the study. **3rd Column:** Result of AR Study.These results show that AR can benefit in both comprehension and information retention.

#### 4.3.2 VR Research in Comprehension and Information Retention

VR learning environments consistently show benefits in areas such as comprehension, memory, and information retention. A study at the University of Warwick in 2018 observed such improvements. Study participants who learned in VR settings demonstrated improved knowledge acquisition and understanding of material compared to those learning via video. VR setting participants also showed better performance in terms of information recollection than those learning via textbook or video. Additionally, participants reported a heightened sense of novelty and interest toward the VR learning medium as a whole, along with a decrease in negative emotions compared to both video and textbook-based options for instruction [8]. This indicates that VR can be helpful for improving comprehension, understanding, and recall of information compared to traditional learning techniques such as textbooks, and the methods used in distance education such as videos.

Another major advantage of using virtual reality to achieve learning objectives is that it is highly motivating. Even before VR became more mainstream, studies with computer based immersive experiences generated positive attitudes towards learning [60]. Mikropoulos et al. created a VR application where future teachers could explore a lake and learn about different sea organisms. After using the application, all of the participants felt the experiences could help in a teaching environment. Furthermore, VR presents novel opportunities to not only improve overall information retention and engagement, but also provides opportunities for learning that cannot be achieved in person. For example, VR is capable of emulating field trips in ways impossible for both distance education as well as in person. In 2019, a field trip-like study by Zhao and Klippel was used to test information retained while in the field [94]. There were 3 groups: a "normal" field trip in person, a basic virtual field trip, and an enhanced virtual field trip. Those in the basic virtual field trip were provided a 10 by 10 area using an HTC Vive HMD, which is a VR HMD. Those in the enhanced virtual field trip used the same setup, and were also provided the option to view locations at an elevated level (27 feet high). All students then undertook the trip from 35-45 minutes. A Spatial Situation Model (SSM) is "a mental representation that comes into play when users attempt to retrieve spatial information from memory" [94]. Given a questionnaire in the style of an SSM, those in the basic virtual field trip performed better than those in the actual field trip, and the enhanced virtual reality performed better than both. It was also observed that those in the basic virtual field trip reported higher mean levels of enjoyment with the field trip than those in the "normal" field trip. This shows that VR can not only emulate a classroom, but can provide valuable educational experiences which could not be provided by an in-person classroom alone [94].

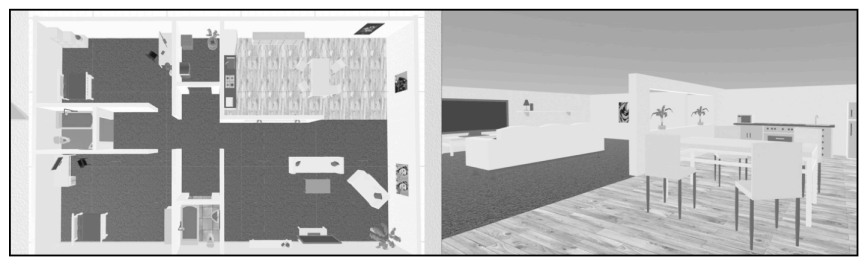

Figure 5: *An Immersive Memory Palace: Supporting the Method of Loci with Virtual Reality*. 2017. Memory Palace simulation used for testing. [41].

While the results of VR and AR studies show that VR is not as significant in promoting comprehension and retention as AR is, the studies show that VR helps with recollection, understanding, and comprehension due to the enhanced realism and visualization compared to traditional online learning (such as the online field trip [94]). For these reasons, it can be useful in education overall. In terms of distance education specifically, the expanding accessibility, ubiquity, and affordability of internet access and VR equipment, as well as the inherent lack of required physical learning environments or classrooms associated with VR, make VR an especially worthwhile consideration for those pursuing distance education.

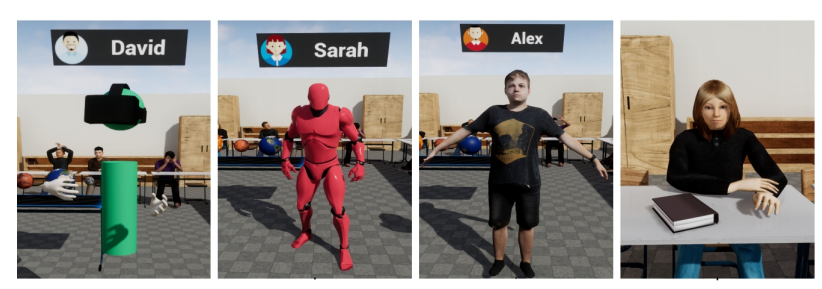

Figure 6: *Scale - Unexplored Opportunities for Immersive Technologies in Place-based Learning*. 2019. From left to right: Models used in Zhao study with increasing levels of realism [94].

| Study | Device | Result |
|---|---|---|
| Allcoat [8] | Head Mounted Display | Improved recollection and engagement |
| Mikropoulos et al. [60] | Head Mounted Display | VR seen as valuable educational tool |
| Zhao [94] | Head Mounted Display | VR helped spatial information retention |

Table 5: *Results of VR Studies to Address Comprehension and Information Retention*. **1st Column:** Reference. **2nd Column:** AR device used for the study. **3rd Column:** Result of AR Study. These results show that VR can help with information recollection.

#### 4.3.3 Future Research in Comprehension and Information Retention

Within AR, limited research has been conducted on information retention for distance learning. While Duenser et al. did test retention for physics education, there were not enough participants to determine conclusively AR's effect on retention [9]. One concern voiced by students in the same study by McBrien et al. mentioned in section 4.2 was that "sometimes it was hard to keep up with the messages, listening to commentators, and reading the lesson," [57] indicating that too much stimuli could harm students' learning experiences. Conducting more research on how different amounts of visual stimuli in an AR or VR experience affect information comprehension would give us a better idea of how to design lessons using VR or AR.

## 5 Instructor Resources in AR/VR Distance Education

As mentioned previously, education AR or VR resources were far too rigid to be utilized by instructors without a lot of coding experience in the past. Many studies have explored the viability of using AR and VR for educational purposes, but these studies used pre-developed content for instructors. Developing content requires advanced skills in both AR and VR technology [50], and is typically created by individuals far removed from the educational subject matter, such as a contracted developer. This makes it very difficult for instructors to customize content for students, as well as adapt and change lesson plans as fluidly as with traditional learning materials [46]. This issue almost entirely prevents teachers from adopting AR or VR resources in education, as this stops educators from designing virtual learning environments that are specific to their material.

#### 5.0.1 Research and Resources for Distance Education in AR

There has been a large push within AR research to facilitate the creation of AR development resources for teachers and laypeople without coding experience. Using the Equator Component Toolkit, a system used to push text-based media to small displays [24], Hampshire et al. was able to create an AR developer application [35]. It features a drag and drop user interface, where developers can use AR content pre-created for general applications. While this has not yet been used in an educational context, it demonstrates the desire for AR creation software, as well as the need for user-specific development tools [35]. Additionally, Seichter's group created ComposAR [77]. ComposAR requires python for added functionality such as interaction or application plugins, but implements a graphic user interface to compile and view code as it is being created.

Specifically for distance education, Ying Li created the AR Environment for Remote Education (ARERE) [91]. Students are able to see and interact with the same virtual object from different locations using this tool. Additionally, the digital objects track physical objects that are distributed to students in advance, so they can tangibly interact with the physical object and observe as it changes digitally. This has not yet been tested on students for feasibility in an actual education environment, but this represents promising potential for AR to give a sense of presence for applications such as distance education.

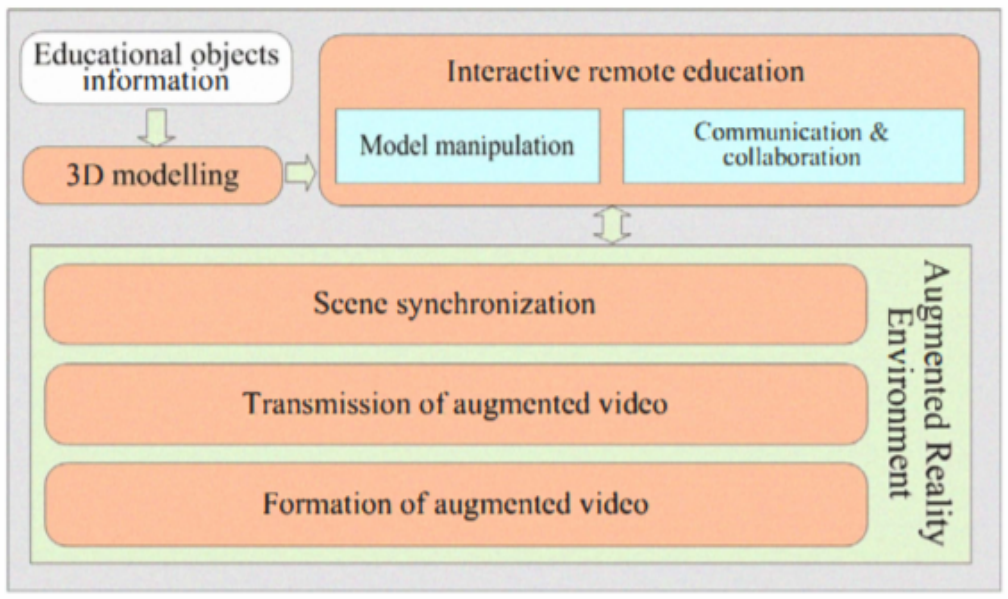

Figure 7: *AR Environment for Remote Education (ARERE) Structure*. [91]

While digital overlays can be extremely useful, it is sometimes necessary for students to be able to physically touch and interact with their learning resources. To provide haptic feedback, instruction materials have utilized real objects with AR markers - easily recognizable 2D symbols such as pictures to designate where to display AR objects - for augmentation through computer vision tracking techniques. This allows students to physically rotate and discover the augmented object, adding sensory feedback to the digital educational experience [29]. In this way, the same mechanical hardware can be used for a variety of teaching levels, reducing hardware cost and increasing content variability. These technologies can greatly aid understanding in areas such as biology, astronomy, mechanics, and other fields where digital overlays can be used to help describe typically unobservable physical structures. Additionally, commercial HMDs such as the HoloLens allow users to interact with digital overlays by touching digital buttons [89]. This, combined with physical objects, can allow for haptic and digital feedback, further aiding the learning process.

Many commercial AR tools also exist to aid instructors in education. Wonderscope offers interactive AR mobile storybooks for children to creatively learn literacy [87]. It features traditional story books such as Little Red Riding Hood, and teaches children how to read. As they read the words, they can watch Little Red Riding Hood perform the actions. ARTutor allows teachers to quickly and easily augment textbooks with AR content [53]. Teachers can upload a book pdf and add 2D overlays, sound, and videos for students. Then, as students read the physical book through the ARTutor smartphone application, they will be able to view the augmented content. As teachers can easily upload any sound, video, or image files, ARTutor provides immense customizability for teachers. Teachers can add audio of themselves explaining a formula in a textbook or a video of a particular battle in a history book. Quiver is another commercially available software that allows users to download educational coloring pages in order to teach students about science and mathematics [71]. The coloring page will then become a 3D image, allowing the student to see their creation come to life by adding a new dimension as well as animations. As the content is not customizable, this application is more suitable for simple subjects with younger children.

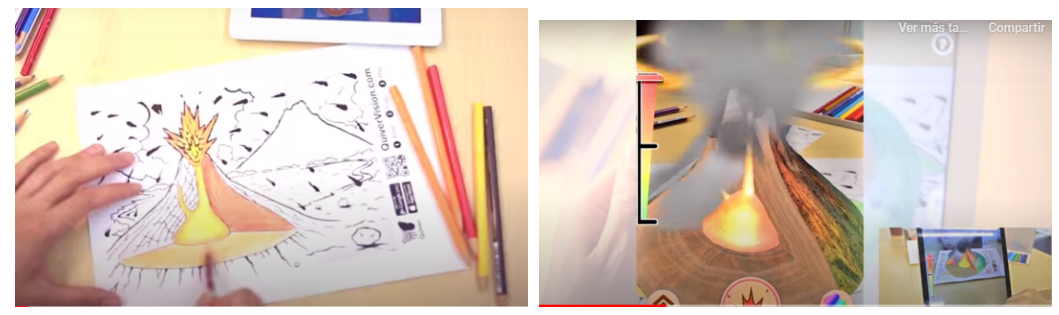

Figure 8: *Demonstration of QuiverEducation for learning about volcanoes.* Students color image of a volcano (left) and watch it erupt through the Quiver smartphone app (right). [71]

#### 5.0.2 Research and Resources for Distance Education in VR

While many studies have explored the viability of using VR for educational purposes, these studies generally used pre-developed content for instructors. One major barrier which still exists with VR or AR content is an easy way to create custom content for these platforms [46, 50]. For this reason, further research into content creation and the ability to customize content for instructors would be helpful in advancing VR educational applications, especially for distance learning. This can include software that allows for creating customized content without intense coding knowledge, as well as dynamic programs, where the teacher can easily adapt current content into a VR learning environment. For distance learning, this could be environments where the teacher can teach normally, and the students will simply experience the content in an immersive way. Youngblut conducted an extensive survey of research and educational uses of virtual reality during the 1990s, and attempted to answer questions about the use and effectiveness of virtual reality in kindergarten through grade 12 education [92]. They found that there are unique capabilities of virtual reality, and the majority of uses included aspects of constructivist learning, which is a theory that that recognizes the learners' understanding and knowledge based on their own experiences prior to entering school. He generally concluded that VR holds potential educational effectiveness for special needs students [92]. However, he also made note of the contexts in which VR is not a suitable option for educational use, and where its downsides are most prevalent, noting that the chief inhibitor of widespread implementation is the lack of customizable content: teachers commonly reported frustration with the lack of ability to customize educational content for a lesson, both before and during the lesson.

#### 5.0.3 Future Research in Immersive Instructional Resources

An increasing amount of AR and VR resources are becoming available to the public, but one of the greatest challenges faced when trying to adopt VR or AR as an educational resource is the inability to easily create virtual environments without programming knowledge. VR and AR are advancing rapidly, so a survey of what resources are currently available, their strengths and weaknesses, and what is needed for them to be ideal for education could be valuable in trying to understand how these platforms can be made more accessible for education. Updated research on how effective new AR and VR educational resources are would give us a direction in how to move forward with developing an AR or VR content creation platform for educators. In order to consider AR or VR as a viable option for any form of education, instructors will need to be able to easily manipulate the virtual environments so that they can work with curricula or lesson plans.Moreover, there is considerable work on developing various tools which might generally be of use in designing effective virtual environments. For example, Rambach et al. [72] discuss the use of VR-based remote collaboration as an alternative for an in-person research conference during the COVID-19 outbreak when large in-person gatherings were not possible. They describe how the conference Laval Virtual utilized VR as part of a remote collaborative system to host the event. In that case, all the participants had virtual avatars, and presenters were able to share their computer screens into the virtual environment. Rambach et al. explain that while simpler tools like basic remote support applications which make use of XR already have commercial applications, larger tools like remote collaboration systems still need to develop in order to be ready for the public's use [72]. Recent developments in terms of tools like VR include note-taking tools [22, 70], which can be useful for education and distance learning. While these tools present interesting possibilities, they also demonstrate that additional work is needed in order for these tools to be ready for broader use and applications. Prior papers [22, 70, 72] also highlight more specific directions on what research needs to be conducted to improve its respective technology.

## 6 Future Research Directions

AR and VR have proven to be valuable educational resources where they fit in, but we have found several issues that need to be overcome in order to make these technologies more widely available to educators. Critiques of traditional distance education include the decreased amount of social interaction, lack of discourse, and absence of non-verbal cues within these environments [23, 44, 57]. VR and AR help remedy these issues, as they encourage communication more effectively than traditional audio or video methods. However, they could still benefit from further advancements. AR HMD's were in the way when two users were trying to communicate and make eye contact [14], so research in how communication is affected with varying types of HMD's would be valuable. VR can effectively facilitate opportunities to socialize, but understanding how successful existing VR educational environments are in supporting discourse is still an area that should be explored. In terms of engagement and focus, more research is needed in determining what amount of visual stimuli helps to engage and focus students without being too distracting.

Distance learning environments have faced criticism for causing disengagement between students and their courses for reasons like technical issues or improper amounts of visual stimulation related to the lesson [57]. Traditional distance educational environments and VR or AR environments will have different requirements of users, and these will likely act as a limitation for widespread adoption of AR and VR. Those limitations are discussed further in sections 4.1.3, 4.2.3, 4.3.3, 5.0.3, and 6. Other factors that negatively impacted student engagement in traditional distance learning environments were related to balancing the amount of stimuli available to the student during the lesson. Understanding what factors are necessary (or unnecessary) for students to have in traditional distance education might give us an idea of how we should be designing virtual environments for students. Additionally, more research is still needed on how virtual lessons affect long-term information retention.

Acceptance of AR or VR technology by teachers or educational systems will also play a role in the widespread adoption of AR or VR in education. Teachers have concerns about the need for continuous training required to be able to use the devices effectively, and opinions about the process of 3D modeling received both positive and negative feedback [83]. Further research in how educators under different age groups feel about immersive technologies would also be valuable in better understanding what teachers think about bringing AR and VR into their classrooms. Studies aimed at trying to gauge teachers' willingness to accept technologies like AR and VR into the classroom do exist, but these studies focus on teachers which are millennials or younger [25,54]. During the 2017-2018 school year, at least 75% of teachers in the United States were 30 years or older, and about a quarter of teachers in the U.S. were over 50 years old [1]. By only studying younger teachers, we are missing out on how a substantial portion of the education workforce might see the adoption of VR or AR into education.

Financial concerns about bringing these devices into the classroom are also important constraints to consider when deciding if AR or VR can be widely used in schools. Immersive devices have gotten cheaper, and devices like Google Cardboard exist now as well. However, robust HMDs may still prove to be challenging to fit into school budgets. Higher-end AR devices like the Microsoft Hololens 2 are also still too expensive to enter the consumer market (the Hololens 2 retails at $3500 USD on the Microsoft Store). Educational systems commonly face funding issues, so more research with regards to whether schools can afford immersive technologies is necessary.

The last major challenge preventing educators from adopting VR or AR into their classrooms is the inflexibility of the technology when it comes to designing lessons. Teachers need to be able to easily create virtual environments related to their curricula without programming knowledge. There have been recent advancements in what resources are available as educational VR or AR tools, so research into what resources are available now, as well as how effective these resources are would be valuable. Many studies have explored the viability of using VR for educational purposes, but these studies generally used pre-developed content for instructors. Current methods are either not customizable, or require much time and practice to create. We had previously mentioned ARTutor which is easy to use and customizable, but it only provides 2D content, and therefore does not take advantage of AR's immersive 3D potential. Work in creating platforms which allow users to create thorough virtual experiences easily would be extremely valuable in pushing VR and AR into education.

A valuable research topic for the future is the establishment of metrics for evaluating immersive environments in education. The methods utilized in this paper were more categorical, and could not numerically quantify what methods for AR and VR are best. Having a universal metric for future researchers to follow will help in evaluating these technologies.

### 6.1 Hardware Advancement

AR and VR hardware are integral to providing an immersive experience for users. Current AR headsets include the HoloLens and digital overlay glasses. However, due to the high cost of the HoloLens and technical limitation of AR glasses, VR headsets with cameras have also been used to create AR experiences. These headsets, along with the HoloLens, are relatively large and quite heavy. This can be uncomfortable, especially for younger students. Further research into increasing computing power or decreasing headset size would be helpful in enhancing the user experience for AR and VR technology. While VR headsets that use controller end-effectors provide better tracking to allow for a more interactive experience, they restrict fine motor movements for the hands, as well as important aids in learning such as handwritten notes, important for distance education. Additionally, a user's depth perception can be negatively impacted when using an HMD, and this is especially prevalent in VR environments. When determining depth of 1.75m (meters) to 7.35m, users had an average absolute error of .21m and .15m for VR and AR respectively [69]. The depth error increases as the distance of an object increases [69], so it's important for developers to display depth-sensitive imagery close to the users.

For VR devices, the amount of motion a user can exhibit is limited to the amount of space accounted for by a device's trackers. Also, low display frame rates, small depth of field and continuous virtual motion without physical motion can create cybersickness [20]. These limitations can create a less immersive or appealing experience for users. For both AR and VR, eye strain and headaches due to having digital content so close to the face is still a issue hindering AR and VR advancement. Research into hardware that can reduce these problems will help the adoption of this technology in education. Additionally, haptic technology is typically limited to simple vibrational tactile feedback. The vibrational feedback is useful, but normal, shear, and kinesthetic feedback can provide a more immersive and interactive experience within learning.

For AR devices, the field of view can be quite poor (about 52 degrees) [90]. Fortunately, Xiong et al. developed a new see-through technology that has the potential to increase the field of view to 100 degrees [90]. Unfortunately, it will be a while before this is incorporated into consumer devices.

### 6.2 Software Advancements

Currently, certain technological aspects of AR and VR hinder their educational advancement. For classrooms, AR and VR technology must become more user friendly from a teacher's perspective [46], so that teachers can utilize the immersive content with similar effort as teaching in traditional methods. From this, future research is recommended in developing more user friendly teaching interfaces, or automating some of the AR or VR content so that the teacher can focus on instruction.

Alternate forms of input are also helpful [39]. This has, of course, been implemented through controllers for VR HMDs and HoloLens, but mobile VR and most AR applications do not utilize hand tracking for interaction. Alternate forms of input increase the interactablility of displays, further engaging students. The HoloLens has the ability for hand and eye tracking [49] and touch interaction [89], and mobile devices can utilize hand tracking to incorporate touch feedback. However, these have not yet been realized or evaluated in an educational context. Utilizing the sense of touch will be very helpful to increase interactivity and engage learners kinetically. From this, future research is recommended in the utilization of haptics and touch interaction in the context of distance education (i.e., note-taking, handwriting, art and drawing, etc.).

Furthermore, educational solutions that leverage AR and VR technologies as a whole must evolve, both in terms of technical feasibility and in acceptance by educators. Like many educational innovations in the past, the use of AR in distance learning classrooms could encounter constraints from schools and resistance among teachers. The learning activities associated with AR and VR usually involve innovative approaches such as participatory simulations and studio-based pedagogy [46]. The nature of these instructional approaches however is quite different from the teacher-centered, delivery-based focus found in conventional teaching methods, where the teacher is the focal point of the student body and student participation is less frequent. Institutional constraints such as covering a certain amount of content within a given time frame also cause difficulties in implementing novel educational initiatives like this [46]. Thus, there may be a gap between the teaching and learning methods currently used in distance learning classrooms and the student-centered and exploratory nature of learning engendered by AR and VR systems. Designers of AR and VR learning environments need to be aware of this gap and provide support to help teachers and students bridge it, ideally in the form of tutorials, listening sessions, and working closely with educators in order to ensure the technology is suitable for the unique challenges education presents.

### 6.3 Scalability Issues

VR and AR do have technical demands that may be impractical for the average present-day classroom. We have seen traditional synchronous distance education courses run into connectivity issues in the university setting where the number of users are large. AR or VR educational settings should make internet connection requirements more demanding than video conferencing software due to the additional data needed for the virtual environments. This means that students will likely need much stronger internet connections to run these devices without interruption, and running a university-like VR lecture hall with over a hundred students might be impossible at the moment because of the internet bandwidth required. However, the development of 5G connection technology could make this possible. 5G promises internet speeds of greater than one gigabit per second [85], which would make streaming large amounts of data much more feasible. Drawbacks to 5G, like possible issues with wall-penetration, do exist, but the bandwidth offered by 5G connectivity should expand on online capabilities for AR or VR. Additional research on stress testing VR and AR online platforms would be helpful for understanding the networking limits of AR and VR systems, as well as how these limits might affect aspects of meaningful education like feelings of presence or meaningful social interactions. While scalability is more of an issue with applciation than one which has to do with research advancements, we need to know the current limits of AR and VR networking systems before research can be done on how to optimize them.

### 6.4 Learning Styles and AR/VR in Education snd Distance Learning

There has been much debate in the attempts to categorize the different ways in which individuals learn. The term "learning styles" has had many definitions over the decades. A general definition of learning styles has been provided by James and Gardner as "the different patterns of how individuals learn." [42] While many categorization methods exist for learning styles, the VARK (Visual, Aural/Auditory, Read/Write, and Kinesthetic) learning styles exist as commonly utilized guidelines by instructors [56]. In terms of educational material: Visual learners prefer to see different formats, space, graphs, charts, diagrams, maps and plans. Aural/auditory learners prefer hearing discussions, stories, guest speakers, podcasts, lesson recordings, and related chat. Read/write learners prefer lists, notes and text in all formats, whether in print or online. Kinesthetic learners prefer senses, practical exercises, cases, trial and error, and examples [56].

The rationale in identifying learning styles in regards to AR and VR education is not only to understand user preference for digital interfaces, but also to explore how AR and VR can be used to curate to different learning styles. However, from what research exists on the impacts of learning styles on AR/VR geared education, it is concluded that a student's learning style is never a hindrance to the effectiveness of AR or VR education [21, 40, 82]. "...While there are many theories, little empirical evidence exists for linking such learning styles to 3D VR education." [40] The reason for this is evidently that there is simply no learning style that is not satisfied by AR/VR-delivered education, as all types of learners benefit from the vast array of applicable AR and VR scenarios for any academic subject.

There exist many examples of AR/VR education being unintentionally curated to VARK learning styles simply by the nature of immersing learners in VR or AR. These include: Visual learning health care students observing a VR-simulated complicated surgical procedure; aurally learning industrial trainees being able to hear ambient noise, specific alerts, or sounds made by equipment that accurately simulate their fast-paced jobsites and keep them engaged in their tasks; read/write oriented learners being able to organize their notes and texts in ways unobstructed by physical limitations; and kinesthetic learners pursuing architecture by being able to use VR and/or AR to explore, manipulate, and adjust scale models, helping create a better understanding of layout, stress points, etc. [2]

Continuing research relevant to learning styles involving the overlap of AR and VR in education seeks to establish methods that can accurately gauge a particular student's learning style and cater a more personalized educational experience tailored to them [52]. This highlights that although learning styles of a given student do not hinder the effectiveness of AR/VR on education, it is possible that capitalizing on learning styles may provide additional benefits.

### 6.5 Future of AR/VR in Education

AR and VR allow digital images to interact in our real world, making them useful in education. This is particularly evident within distance education, where a teacher cannot always be present with a student. Through the studies discussed in section 4, AR and VR have been shown to increase social interaction [14, 62, 67, 73, 75, 79], comprehension [8, 9, 27, 37, 50, 60, 76, 94], and engagement [28, 38, 43, 47]. More research in AR/VR specifically in distance learning would be beneficial so as to better solidify these results. Additionally, multiple hindrances such as technology and hardware limitations as well as cost impede their adoption in education institutions. While studies suggest AR and VR can be most beneficial in K-12 education, where interaction, engagement, and focus play a larger role in the learning process, the cost of hardware, educator adoptions, and developing content for such a wide variety of curriculum reduce the practicality of adoption within K-12 schools. Currently, it seems as if both AR and VR are great supplemental resources for teachers looking to explore either specific subjects which people have built immersive resources for, or more common/better known topics which multiple forms of content likely exists for. They cannot currently cover everything a student must learn in K-12 education, but they can help to reinforce knowledge and keep students engaged.

## 7 Conclusion

AR and VR are prime candidates for addressing the various educational challenges presented by distance learning, especially during the COVID-19 pandemic, but we find they are currently inadequate for widespread adoption in the educational community and require further technological developments to be truly successful in the context of education. The COVID-19 pandemic has heightened the need for interactive and engaging educational environments, as the forced transition to distance learning has resulted in instruction being generally limited to discussions and lectures presented via video conferencing tools such as Zoom. This widespread implementation of distance learning has led to decreased student motivation and information retention, primarily due to the fact that distance learning instruction styles are often less engaging than if the content was delivered in-person.

At first glance, and especially in terms of individual features, AR and VR educational solutions appear to be ideal choices to help combat these challenges with student motivation, engagement, and information retention. Due to the fact they closely mimic learning in-person through a 3-dimensional environment, the use of AR and VR technologies as educational tools has been demonstrated to improve knowledge acquisition among a variety of educational subjects as compared to traditional distance education (and, in certain cases, as compared to traditional in-person education). Furthermore, AR and VR have also been shown to improve student engagement and motivation in the classroom, as immersive lectures are much more interesting and comprehensive than lectures presented on a 2-dimensional screen. AR and VR educational solutions were already being used before the pandemic for these reasons, in contexts such as supplementing in-person instruction through an AR-based anatomy lesson.

Despite presenting numerous advantages over both in-person instruction and distance learning, AR and VR educational solutions are still not technically mature enough yet to be implemented in a widespread fashion as a direct solution to many of the challenges presented by distance learning. In this paper, we note how a chief inhibitor of widespread adoption by the educational community is the lack of customizable content: teachers commonly reported frustration with the lack of ability to customize educational content for a lesson, both before and during the lesson. Additionally, further research to reduce motion sickness and eye strain for AR and VR is necessary before incorporating these devices into a required curriculum, as students must feel comfortable learning for long periods of time without fear of negatively impacting their health. Finally, the cost of high-end versions of AR and VR devices is also a major inhibitor of their widespread use: educators must be able to buy these devices in bulk for affordable prices. We hope that this paper brings to light these limitations and aids researchers in assessing the current state of immersive technologies as educational resources. We suggest areas of improvement to promote adoption of AR and VR in educational environments for distance learning, with the goal that AR and VR may soon be able to supplement or even completely replace the entire in-person classroom experience.